\documentclass[%
  reprint,
  superscriptaddress,
  longbibliography,
  preprintnumbers,
  amsmath,
  amssymb,
  aps,
  pra,
  floatfix,
  tightenlines
]{revtex4-2}

\usepackage{caption}
\usepackage{changes}
\usepackage{float}
\usepackage{physics}
\usepackage{booktabs}
\usepackage{graphicx}
\usepackage{dcolumn}
\usepackage{bm}
\usepackage[percent]{overpic}
\usepackage{changes}
\usepackage{cancel}
\usepackage{subcaption}
\usepackage[absolute,overlay]{textpos}

\usepackage[colorlinks=true,
citecolor=green,
linkcolor=purple,
anchorcolor=black,
urlcolor=purple]{hyperref}

\begin{document}


\title{Tomography-assisted noisy quantum circuit simulator\\ using matrix product density operators} 

\author{Wei-guo Ma}
\affiliation{Institute of Physics, Chinese Academy of Sciences, Beijing 100190, China}
\affiliation{School of Physical Sciences, University of Chinese Academy of Sciences, Beijing 100049, China}

\author{Yun-Hao Shi}
\affiliation{Institute of Physics, Chinese Academy of Sciences, Beijing 100190, China}
\affiliation{School of Physical Sciences, University of Chinese Academy of Sciences, Beijing 100049, China}

\author{Kai Xu}
\email{kaixu@iphy.ac.cn}
\affiliation{Institute of Physics, Chinese Academy of Sciences, Beijing 100190, China}
\affiliation{School of Physical Sciences, University of Chinese Academy of Sciences, Beijing 100049, China}
\affiliation{Beijing Academy of Quantum Information Sciences, Beijing 100193, China}
\affiliation{CAS Center for Excellence in Topological Quantum Computation, UCAS, Beijing 100049, China}

\author{Heng Fan}
\email{hfan@iphy.ac.cn}
\affiliation{Institute of Physics, Chinese Academy of Sciences, Beijing 100190, China}
\affiliation{School of Physical Sciences, University of Chinese Academy of Sciences, Beijing 100049, China}
\affiliation{Beijing Academy of Quantum Information Sciences, Beijing 100193, China}
\affiliation{CAS Center for Excellence in Topological Quantum Computation, UCAS, Beijing 100049, China}


\begin{abstract}
  In recent years, efficient quantum circuit simulations incorporating ideal noise assumptions have relied on tensor network simulators, particularly leveraging the matrix product density operator (MPDO) framework. However, experiments on real noisy intermediate-scale quantum (NISQ) devices often involve complex noise profiles, encompassing uncontrollable elements and instrument-specific effects such as crosstalk. To address these challenges, we employ quantum process tomography (QPT) techniques to directly capture the operational characteristics of the experimental setup and integrate them into numerical simulations using MPDOs. Our QPT-assisted MPDO simulator is then applied to explore a variational approach for generating noisy entangled states, comparing the results with standard noise numerical simulations and demonstrations conducted on the Quafu cloud quantum computation platform. Additionally, we investigate noisy MaxCut problems, as well as the effects of crosstalk and noise truncation. Our results provide valuable insights into the impact of noise on NISQ devices and lay the foundation for enhanced design and assessment of quantum algorithms in complex noise environments.

\end{abstract}

\maketitle


\section{\label{sec:level1}Introduction}

Simulating quantum circuits on classic computers in the presence of noise represents a formidable challenge, crucial for advancing quantum computing, especially with noisy intermediate scale quantum (NISQ) devices~\cite{Preskill2018quantumcomputingin, cheng_noisy_2023}. The inherent difficulties of simulating random quantum circuits without noise due to system size and circuit depth have been well-documented~\cite{Aaronson2005Quantum, Bremner2011ClassicalSimulation, PhysRevLett.117.080501}. To reduce these complexities, research has shifted towards approximate simulation methods, such as those employing matrix product states (MPS), which have been instrumental in simulating realistic quantum dynamics~\cite{PhysRevX.10.041038, PhysRevLett.93.040502, SCHOLLWOCK201196, fannes_finitely_1992, doi:10.7566/JPSJ.91.062001}. However, MPS struggles to represent mixed states prevalent in noisy quantum systems, prompting the adoption of matrix product operator (MPO)~\cite{Pirvu_2010, PhysRevLett.93.207204, Noh2020efficientclassical} and matrix product density operator (MPDO) frameworks. The MPDO approach, in particular, offers a robust framework for simulating quantum circuits with noise, providing a more accurate representation of noisy quantum systems \cite{cheng_simulating_2021}.

Our research is based on the exploration of tensor network techniques for simulating quantum dynamics, where the limitations of entanglement in classical simulations have led to significant insights~\cite{PhysRevLett.131.180601, doi:10.1137/050644756, terhal_adaptive_2004}. This foundation supports our approach, utilizing the MPDO framework and QPT~\cite{nielsen2000quantum, torlai_quantum_2023} to better capture the intricate behavior of quantum systems in the presence of real noise. The evolving landscape of quantum computing, with the advent of NISQ devices, underscores the importance of our work, as these devices offer a practical context for evaluating the effectiveness of quantum algorithms in real-world scenarios~\cite{bharti_noisy_2022, huang_near-term_2023, lee_evaluating_2023, RMP_McArdle2020, RMP_Cai2023}. Furthermore, tensor network simulations have provided a deeper understanding of quantum dynamics, enriching our approach to include comprehensive noise characterization and simulation~\cite{Jaschke_2019, PhysRevLett.93.207205, PhysRevLett.114.220601}.

In this paper, we present a methodological advancement by integrating QPT with the MPDO framework to achieve a more accurate simulation of noisy quantum circuits. This integration allows for a detailed characterization of operational dynamics and the incorporation of experimental noise data, thereby enhancing the fidelity and practicality of our simulations. Furthermore, the flexibility of our approach allows researchers to introduce experimental data featuring varying levels of noise, contributing to a comprehensive investigation of the impact of noise on quantum simulations. This capability is particularly valuable as it not only enhances the reliability of simulation outcomes but also broadens the scope for collaborative research across different quantum computing platforms. Moreover, by allowing for the dynamic integration of real-world experimental noise into the simulation framework, our methodology paves the way for a deeper and more nuanced understanding of the practical constraints and opportunities present in quantum computing.

This paper is organized as follows: Section~\ref{sec2} introduces the theoretical background and the necessary preliminaries for understanding MPDOs. Section~\ref{sec3} details the methodology, elaborating on the integration of experimental noise data into our MPDO-based simulation framework. Section~\ref{sec4} presents our numerical simulation setup and the results of our fidelity analysis under various noise conditions. Section~\ref{sec5} summarizes our main results and discusses the potential prospects.

\section{Preliminary}
\label{sec2}
\subsection{Quantum Noise Representation}
In the realm of quantum information theory, general quantum noise channels are represented through various mathematically equivalent frameworks. All these representations share the attribute of being completely positive and trace-preserving (CPTP) and effectively capture the impact of noise on quantum states and operations, such as Kraus operators, Choi matrices, and superoperators~\cite{nielsen2000quantum, Kraus1983StatesEffectsOperations, CHOI1975285, Bengtsson_Zyczkowski_2006}.

A widely recognized approach for characterizing the noise involves the use of Kraus representation, which can be expressed as
\begin{equation}
  \mathcal{U}(\rho) = \sum_i K_i\rho K_i^\dagger,
  \label{eq_kraus}
\end{equation}
where the Kraus operator $K_i$ satisfies the completeness condition $\sum_iK_i^\dagger K_i=I$. Each noise channel $K$ consists of multiple operation elements $K_i$, each of which distinct influences on the quantum system.

A noise model may encompass different types of noise, such as the depolarizing noise model of a single qubit. In this model, the qubit may flip with an error rate of $p$. Additionally, it may experience a bit flip or a phase flip with equal probability. This impact on the density matrix can be modeled as
\begin{equation}
  \begin{aligned}
    \mathcal{U}_{\mathrm{DC}}(\rho) &= (1-p)\rho + \frac{pI}{2}\\
    &= (1-p)\rho + \frac{p}{3}(X\rho X + Y\rho Y + Z\rho Z),
  \end{aligned}
\end{equation}
where $X$, $Y$, and $Z$ represent the Pauli-X, Pauli-Y, and Pauli-Z operators, respectively, and $p$ is the probability of depolarizing noise occurrence.

The Kraus representation of the depolarizing noise channel, incorporating the operation elements, can be expressed as follows~\cite{nielsen2000quantum}:
\begin{equation}
  \begin{aligned}
    &K_0 = \sqrt{1 - \frac{3p}{4}}
    \left[
      \begin{array}{cc}
        1 & 0\\
        0 & 1
      \end{array}
    \right],\quad
    K_1 = \sqrt{\frac{p}{4}}
    \left[
      \begin{array}{cc}
        0 & 1\\
        1 & 0
      \end{array}
    \right],\\
    &K_2 = \sqrt{\frac{p}{4}}
    \left[
      \begin{array}{cc}
        0 & -i\\
        i & 0
      \end{array}
    \right],\qquad\ \ 
    K_3 = \sqrt{\frac{p}{4}}
    \left[
      \begin{array}{cc}
        1 & 0\\
        0 & -1
      \end{array}
    \right].
  \end{aligned}
\end{equation}

Thus, the depolarizing noise on a single-qubit gate with representation as Eq.~\!(\ref{eq_kraus}) can be written as
\begin{equation}
  \mathcal{U}_{\text{DC}}(\rho) = K_0\rho K_0^\dagger + K_1\rho K_1^\dagger + K_2\rho K_2^\dagger + K_3\rho K_3^\dagger.
\end{equation}

\subsection{Noise in Circuit Model}
Quantum circuits are inherently susceptible to various forms of noise that can significantly affect their performance. It is crucial to set up models that simplify the representation of quantum noise and help manage this complexity. One such model, the unified noisy circuit model ~\cite{georgopoulos_modelling_2021}, begins with two qubits in the state $\ket{00}$. These two qubits undergo a series of processes, including the state preparation and measurement channel (SPAM) and the thermal relaxation channel (TRC)~\cite{nielsen2000quantum}, which introduce errors over time. Notably, the impact of TRC is time-dependent, with its error rate often being influenced by the duration of the quantum gate process, particularly when compared to the overall circuit duration.

\begin{figure*}[htbp]
  \centering
  \includegraphics[width=0.8\linewidth]{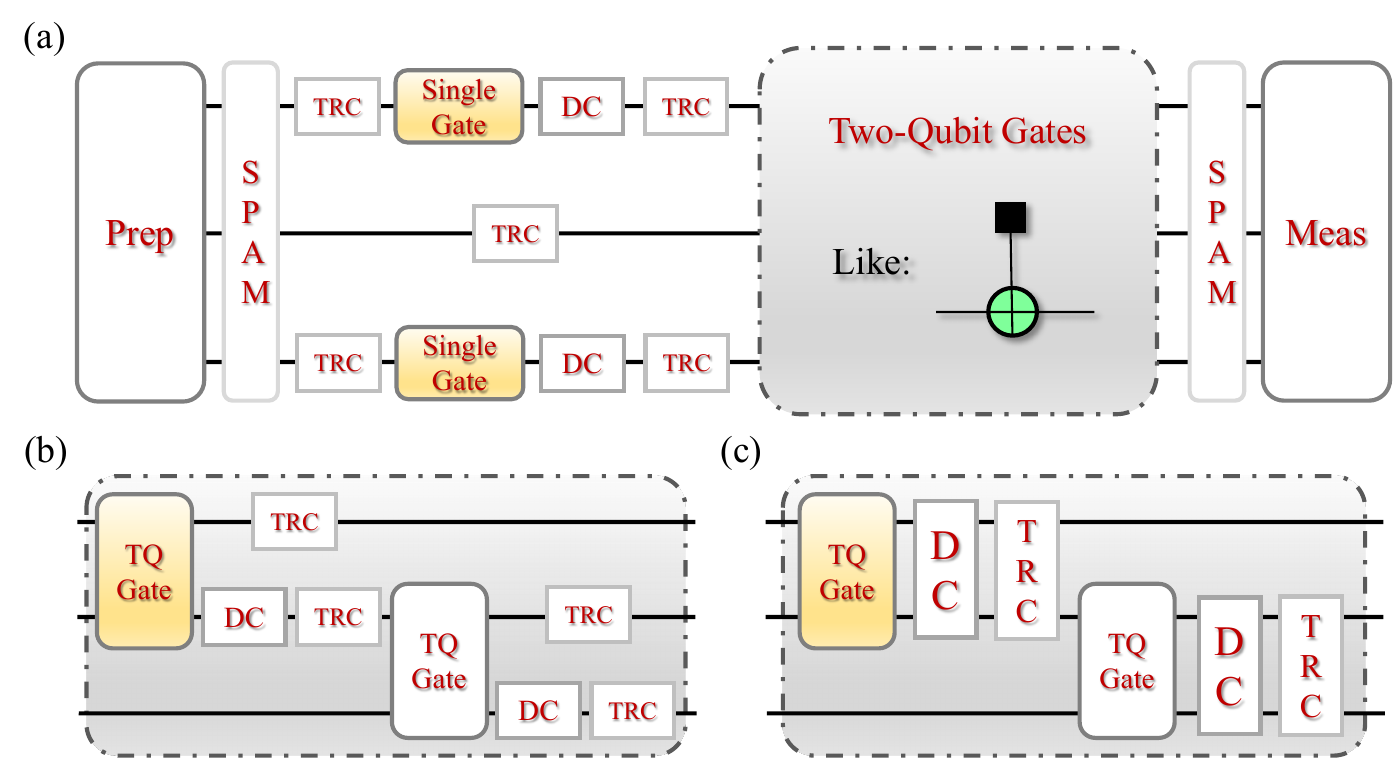}
  \caption{Quantum noise model. (a) The general quantum noise arrangement in an actual circuit involves noise channels with time evolution and quantum operations. Errors can occur during the state preparation phase, propagate during single-qubit gates, and accumulate over time through processes like depolarizing and thermal relaxation channels. The effect of noise in two-qubit gates varies across different models, with efficient methods devised to simulate and address these noise sources. Errors may also manifest during the measurement stage. (b) The unified noisy circuit model. In this model, the depolarizing channel is exclusively applied to the `target' qubit, sparing the `control' qubit from this particular error source. (c) General ideal noisy circuit model. In contrast to the unified model, this ideal model imposes both depolarizing and thermal relaxation channels on the qubits where two-qubit gates are applied, resulting in more complex noise interactions.}
  \label{fig_unifiedNoise}
\end{figure*}

After the application of quantum gates, the qubits encounter the Depolarizing Channel (DC), which, in scenarios involving two-qubit gates, affects only the target qubit, as shown in Fig.~\!\ref{fig_unifiedNoise}(b). Before measurement, additional SPAM process will bring further errors with a probability $p_M$. An alternate model, illustrated in Fig.~\!\ref{fig_unifiedNoise}(c), considers both the depolarizing and thermal relaxation channels simultaneously acting on the control and target qubits. This leads to a more intricate interaction of noise within qubits.

These models provide a framework for simulating the effects of noise on quantum circuits. However, capturing the full complexity of noise experienced in practical quantum computing, including crosstalk~\cite{PRXQuantum.3.020301, PhysRevLett.126.230502} and multiqubit depolarization~\cite{ZHANG2013136, siomau_entanglement_2010}, remains challenging. The QPT method offers a promising approach to represent these complex noise profiles and integrate them into quantum circuit simulations, which is achieved through the combination of an MPDO-based simulator.

\subsection{Noise in MPDO Framework}
In contrast to ideal quantum operators in pure quantum systems, a noisy quantum operator is a higher-order tensor with an additional `leg' connected to its conjugate space operator to express the sum of noise. The traditional MPS framework cannot effectively express the density matrix of a noisy quantum state, i.e., the existence of the additional `leg'. However, this can be effectively portrayed in the MPDO form~\cite{PhysRevLett.93.207204}. Recent work has demonstrated the effectiveness and accuracy of the MPDO method for simulating quantum circuits~\cite{cheng_simulating_2021}.

\begin{figure}[htbp]
  \centering
  \includegraphics[scale=0.55]{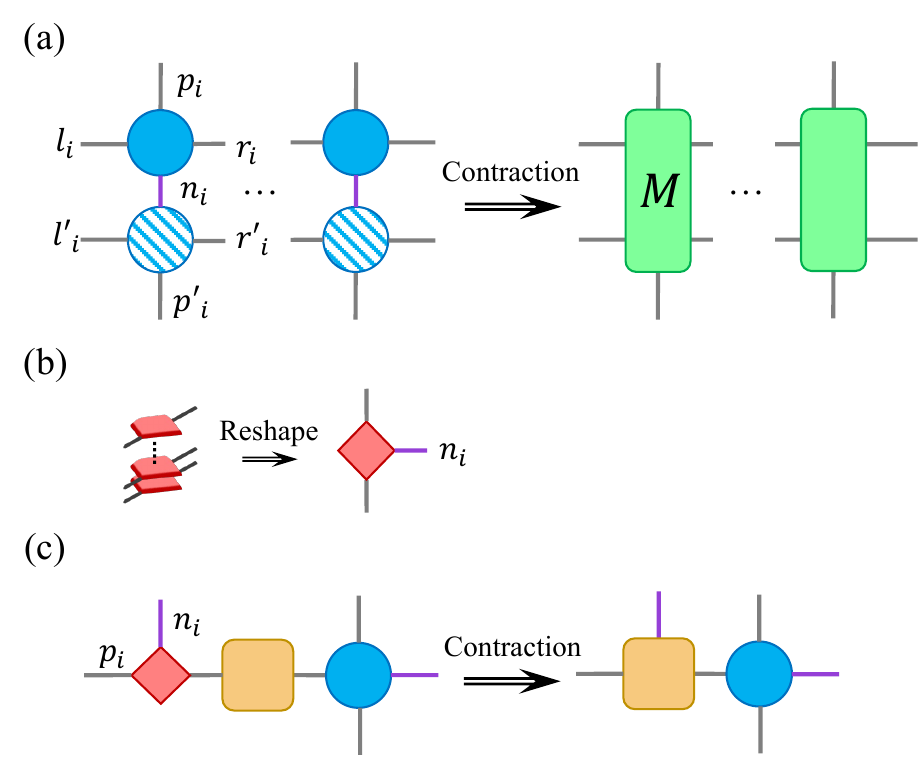}
  \caption{(a) MPDOs representation of the density matrix for an open quantum system, where qubits are linked to their conjugate spaces by a purple leg, introduced by quantum noise. Each qubit is modeled by a blue tensor denoted as $T^{[b]}$, while its corresponding conjugate is represented by $T^{[p]}$. Index $i$ represents the content associated with the $i$-th bit. Through the contraction process, $T^{[b]}$ and $T^{[p]}$ yield the density matrix $M$. (b) Transforming the Kraus representation of a quantum noise channel into a higher-order tensor representation. (c) Implementing the high-order tensor representation of the noise channel on a single-qubit quantum gate. Upon contraction, it gives rise to a noisy single-qubit gate, which is then connected to the respective qubit.}
  \label{fig_mpdoSingle}
\end{figure}

In the MPDO method, a qubit is represented by a rank-4 tensor
\begin{equation}
  T^{[b]} = T_{l_i, r_i}^{p_i n_i},
\end{equation}
where $i$ represents the $i$-th qubit in a qubit system, and $p_i$ refers to the physical index of the qubit. The indices $l_i$ and $r_i$ correspond to the left and right inner bond indices, arising from multiqubit quantum gates that create quantum entanglement between qubits. The index $n_i$ characterizes the noise inner index, facilitating the connection between a noisy qubit and its conjugate copy, thereby representing classical information entanglement. The qubit is depicted as the blue tensor in Fig.~\!\ref{fig_mpdoSingle}(a).

Its conjugate tensor (diagonal stripes) is represented as $T^{[p]} = T_{l'_i, r'_i}^{p'_i n_i}$, where $T^{[p]}_{\quad :, p_i, n_i, :} = (T^{[b]}_{\quad :, n_i, p_i, :})^*$; the superscripts $[b]$ and $[p]$ are used to distinguish the $\ket{\cdot}$ and $\bra{\cdot}$ spaces. Thus, the density matrix for a local qubit can be modeled as
\begin{equation}
  M_{L_i, R_i}^{p_i, p'_i} = \sum_{n_i = 0}^{d^i-1}T_{l_i, r_i}^{p_i n_i}(T_{l'_i, r'_i}^{p'_i n_i})^*,
\end{equation}
where $d_i$ denotes to the dimension of the $i$-th qubit's noise bond.

Consequently, the density matrix describing an open quantum system with $N$ qubits is represented as
\begin{equation}
  \begin{aligned}
    \rho = \sum_{p_0, p'_0, \dots, p_{N-1}, p'_{N-1} = 0}^1 &\mathrm{Tr}\Large(M_{R_0}^{p_0, p'_0}I^{R_0, L_1}M_{L_1R_1}^{p_1, p'_1}\\
    &\cdots I^{R_{N-2}, L_{N-1}}M_{L_{N-1}}^{p_{N-1}, p'_{N-1}}\Large)\\
    &\ \ \dyad{p_0, \dots, p_{N-1}}{p'_0, \dots, p'_{N-1}},
  \end{aligned}
\end{equation}
where $L_i = l_i \otimes l'_i$ and $R_i = r_i \otimes r'_i$.

A noisy Kraus operator element $K_i$ is composed of the noise operation $\mathcal{N}_i$ and the ideal quantum operator $\tilde{K}$,
\begin{equation}
  K_i = \mathcal{N}_i\tilde{K},
\end{equation}
where $\mathcal{N}_i$ and $\tilde{K}$ are both rank-2 tensors for a single qubit. As shown in Eq.~\!(\ref{eq_kraus}), the noise operator elements are applied in parallel to quantum operators, meaning that each element $K_i$ is applied independently to the density matrix $\rho$, followed by a summation of the index $i$. Consequently, these rank-2 tensors, which are noise operations, can combine to form a rank-3 tensor, as illustrated in Fig.~\!\ref{fig_mpdoSingle}(b), wherein the purple leg represents the new bond $n_i$, referred to as the noise bond. The noise node is interconnected single-qubit with the quantum gate node, reflecting the process depicted in Fig.~\!\ref{fig_mpdoSingle}(c). This implies that various types of noise can be applied to a quantum gate. Through this connection, the noise node can be contracted into the quantum gate, thereby generating the noisy quantum gate node with an additional noise bond represented in purple. It is easy to generalize this to multiqubit gates.

\subsection{Low-rank Matrix Approximation}
The utilization of singular value decomposition (SVD) represents a prominent advantage when employing tensor networks to simulate quantum circuits. The SVD technique allows for the truncation of bond dimensions, leading to a reduction in the complexity of the system while maintaining a good approximation of the original system.

To effectively control the noise bond dimension for given qubits, the corresponding tensors can be expressed in SVD form as follows
\begin{equation}
  T_{l_i, r_i}^{p_i n_i} \simeq \sum_\nu U_{l_i, r_i}^{p_i \nu}\Sigma_\nu V_{\nu, n_i}^\dagger,
  \label{eq_svdNoise}
\end{equation}
where $S_\nu$ represents the singular value of the original tensor $T$, arranged in descending order. The matrices $U$ and $V$ correspond to unitary matrices, satisfying $UU^\dagger = I$ and $VV^\dagger = I$, respectively.

\begin{figure}[htbp]
  \includegraphics[scale=0.5]{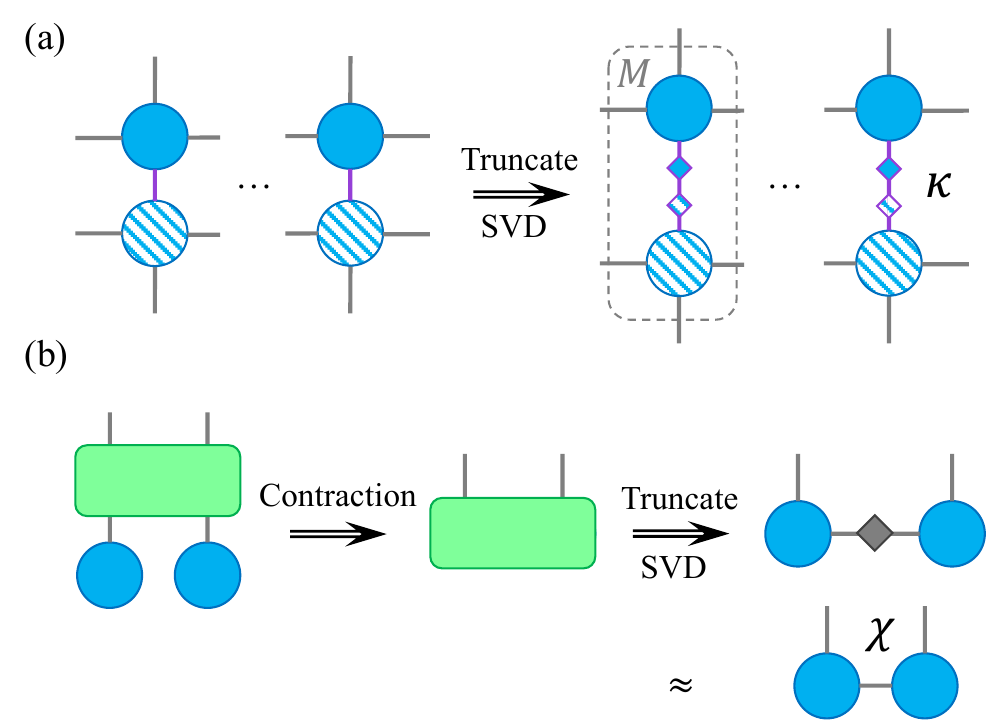}
  \caption{(a) SVD truncation on the dimension of the noise inner indices with limitation $\kappa$. (b) SVD truncation on the dimension of the bond indices with the limitation $\chi$.}
  \label{fig_svdTruncate}
\end{figure}

The above decomposition holds when the dimension limit is greater than or equal to the actual dimension of the original tensor. Figure~\ref{fig_svdTruncate}(a) demonstrates the process of SVD truncation for noise inner bond of qubits and its conjugate copy. By truncating the $\nu$ dimensions in Eq.~\!(\ref{eq_svdNoise}), we obtain an approximation error denoted by
\begin{equation}
  \begin{aligned}
    \left\| T - T' \right\|_F = \min_{\mathrm{rank}(T)\leq \kappa}\left\| T - T' \right\|_F = \sqrt{\lambda_{\kappa + 1}^2 + \cdots + \lambda_{m}^2},
  \end{aligned}
\end{equation}
where $T$ is the original tensor, $T' = U\Sigma V^\top$, and $\lambda_i$ represents the eigenvalues arranged in descending order~\cite{tropp2023randomized}.

By limiting the noise bond dimension through truncation, even a few bond dimensions can accurately represent a significant noise signal. The density matrix of the local qubit thus can be expressed as
\begin{equation}
  \begin{aligned}
    M &\approx \sum_{\nu=1}^{\kappa} \sum_{\nu'=1}^\kappa U_{l_i,r_i}^{p_i,\nu} \Sigma_\nu V_{\nu,n_i}^\dagger \delta_{\nu\nu'} \left(U_{l_i',r_i'}^{p_i',\nu'} \Sigma_{\nu'} V_{\nu',n_i}^\dagger\right)^\dagger\\
    &=\sum_{\nu = 1}^\kappa U_{l_i, r_i}^{p_i \nu}\Sigma_\nu V_{\nu, n_i}^\dagger(U_{l_i', r_i'}^{p_i' \nu}\Sigma_\nu V_{\nu, n_i}^\dagger)^\dagger\\
    &= \sum_{\nu = 1}^\kappa U' \Sigma^2 U'^{\dagger},
  \end{aligned}
\end{equation}
where $U'$ is the truncated $U$ matrix.

The process illustrated in Fig.~\ref{fig_svdTruncate}(b) is similar to the previous one, but now it is specifically employed to truncate the dimension of the bond between qubits. To achieve a unique and globally optimal SVD truncation, a series of QR decomposition from left to right is necessary to transform the matrix product operator into a canonical form (without the conjugate copy)~\cite{cheng_simulating_2021, PhysRevB.78.155117}. After the canonicalization process, the SVD is applied to truncate each of the bond indices from right to left.

\begin{figure*}[htbp]
  \centering
  \includegraphics[width=\linewidth]{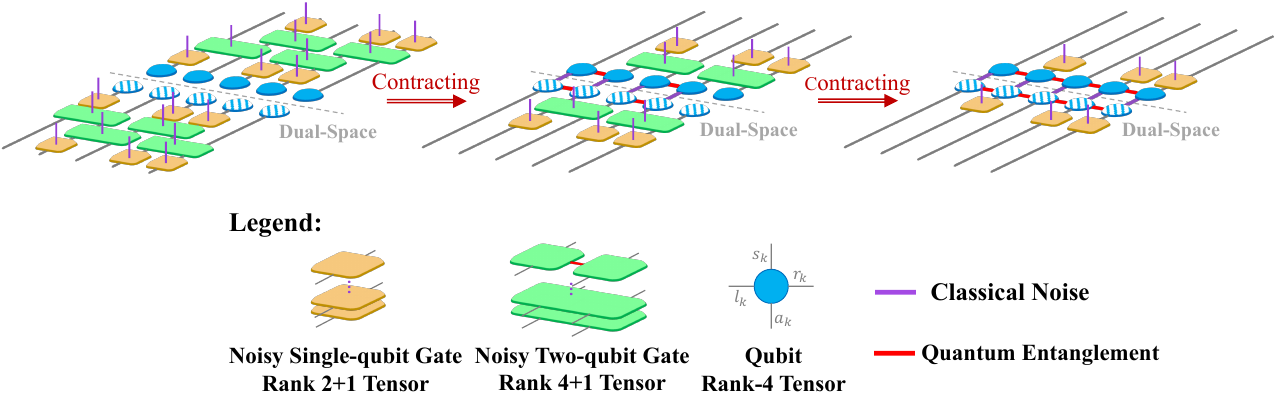}
  \caption{Quantum circuit contraction for MPDOs. Purple lines represent classical noise, and red lines depict quantum entanglement between qubits. The application of noisy gates introduces purple noise lines, and two-bit gates establish red entanglement lines between the qubits.}
  \label{fig_contractionScheme}
\end{figure*}

By implementing this truncation strategy, the calculation complexity of contracting tensor networks is significantly and efficiently reduced. Therefore, the SVD and QR decomposition help reduce bond dimensions and computational resources required for tensor network contractions, making it an advantageous approach for various applications~\cite{paredes2003incommensurate}. After the truncation, MPDO quantum circuit can contract efficiently in the scheme shown in Fig.~\!\ref{fig_contractionScheme}.

\section{Real Noise from Experiment}
\label{sec3}
In practical quantum computing experiments, it is imperative to take into account a multitude of noise sources that extend beyond the inherent quantum noise illustrated in Fig.~\!\ref{fig_unifiedNoise}.~These additional sources include crosstalk, multiqubit depolarization, sparasitic coupling~\cite{PhysRevX.11.021058}, and various forms of noise arising from imperfections in quantum hardware and environmental influences. These complex noise phenomena cannot be readily encapsulated, they cast a significant impact on the performance of quantum circuits and the fidelity of quantum computations, akin to dark clouds looming over the quantum landscape.

\begin{figure}[h]
  \includegraphics[width=\linewidth]{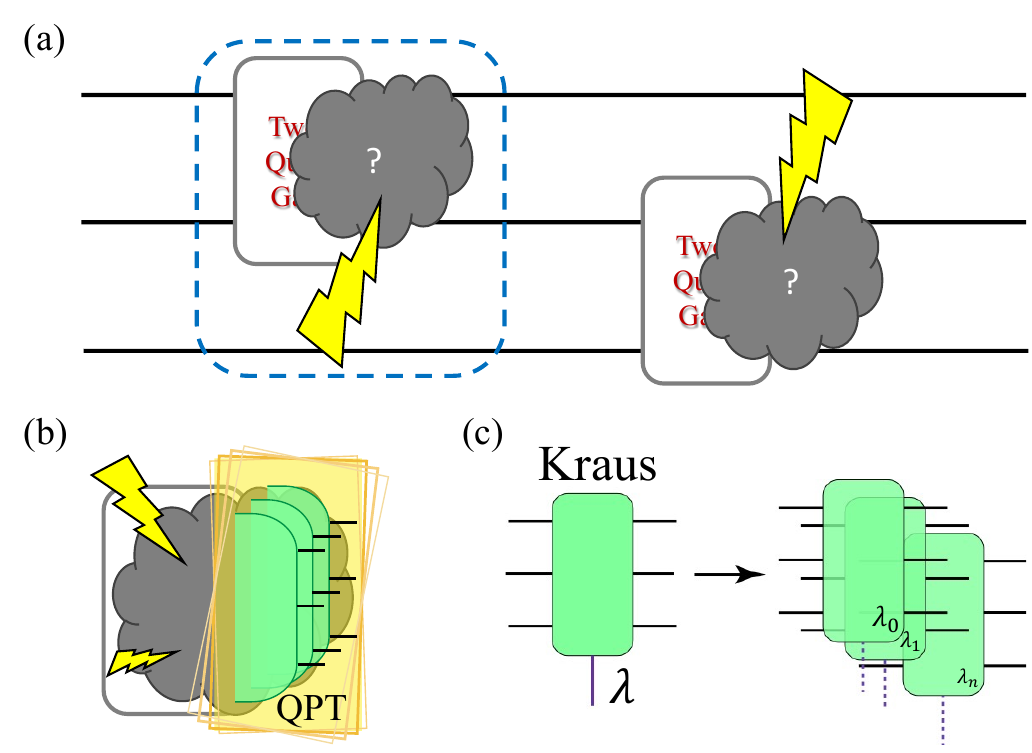}
  \caption{Experimental quantum circuit. (a)  Illustration depicting a quantum gate operation, where unaccounted environmental noises are represented as a dark, nebulous cloud surrounding the gate. This emphasizes the ubiquitous presence of noise during gate execution, affecting the fidelity of the quantum operation. (b) The application of quantum process tomography to a noisy two-qubit gate, specifically highlighting the phenomenon of crosstalk. The crosstalk effect, depicted as an interaction with the nearest qubit, is crucial for understanding and characterizing the intricate noise dynamics in multiqubit systems. (c) Utilizing singular value decomposition to extract the dominant noise tensor from the gate tensor.}
  \label{expNoise_fig}
\end{figure}

Advanced QPT techniques~\cite{nielsen2000quantum, torlai_quantum_2023}, enable us to gain comprehensive insights into the operations of noisy quantum circuits or gates, as illustrated in Fig.~\!\ref{expNoise_fig}(b). By employing this method, we can dissect noisy gate functions into a series of Kraus operators. This process involves eigenvalue decomposition and strategic pruning of these tensors, organized by descending eigenvalue magnitude, allowing for the effective integration of real noise in circuit simulations. QPT provides a powerful tool for researchers to decode the dynamics of quantum processes using experimental data. It facilitates the reconstruction of quantum channels or operations executed by quantum circuits.

\begin{figure*}[t]
  \centering
  \includegraphics[width=\linewidth]{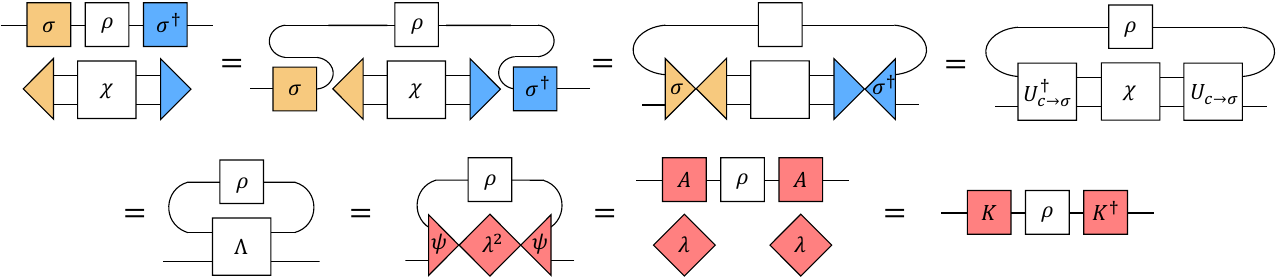}
  \caption{Illustration of the conversion of the $\chi$ matrix to Kraus operators within a tensor network framework. The process begins with transforming the $\chi$ matrix into the Choi matrix via a basis transformation. Following this, an eigenvalue decomposition is performed on the Choi matrix, resulting in the derivation of Kraus operators. The different colors represent various summation operations involved in this transformation process.}
  \label{fig_chi2kraus}
\end{figure*}

It is crucial in reconstructing quantum channels or operations, elucidating the underlying mechanics of quantum circuits. Despite its utility, QPT primarily yields the quantum process matrix (or $\chi$ matrix), represented as
\begin{equation}
  \varepsilon(\rho) = \sum_{mn} \chi_{mn}\sigma_m \rho \sigma_n^\dagger.
  \label{eq_chi}
\end{equation}

To integrate these findings into a quantum circuit, it is necessary to transform the $\chi$ matrix into Kraus operators. This transformation begins by converting the $\chi$ matrix into the Choi matrix, starting with Eq.~\!\ref{eq_chi}. The $\chi$ matrix, defined with respect to an orthonormal operator basis $\{\sigma_m\}$, is related to the Choi matrix $\Lambda$ through the change of basis
\begin{equation}
  \chi = U_{c\rightarrow\sigma}\Lambda U_{c\rightarrow\sigma}^\dagger,
\end{equation}
where $U_{c\rightarrow\sigma}$ is the vectorization change of basis operator. Thus, we have
\begin{equation}
  \Lambda = \sum_{mn}\chi_{mn}|\sigma_m\rangle\rangle\langle\langle\sigma_n|,
\end{equation}
with $|\cdot\rangle\rangle$ denoting the vectorization of an operator. Next, an eigenvalue decomposition is performed on the Choi matrix
\begin{equation}
  \Lambda = \sum_i\eta_i\ket{\psi_i}\bra{\psi_i},
\end{equation}
where $\eta_i$ are the eigenvalues and $\ket{\psi_i}$ are the corresponding eigenvectors. The Kraus operators are then derived from the eigenvectors of the Choi matrix
\begin{equation}
  K_i = \lambda_iA_i,
\end{equation}
where $\lambda_i = \sqrt{\eta_i}$ and $A_i$ is the operator satisfying $\ket{A_i}\rangle=\ket{\psi_i}$. The number of Kraus operators is equal to the rank of the Choi matrix. This process is elucidated in Fig.~\!\ref{fig_chi2kraus} within a tensor network framework. Once real noisy quantum gate data are acquired in the form of Kraus operators, it is integrated into a noisy quantum gate using the MPDO quantum circuit simulator.

\begin{figure}[H]
  \centering
  \includegraphics[width=0.7\linewidth]{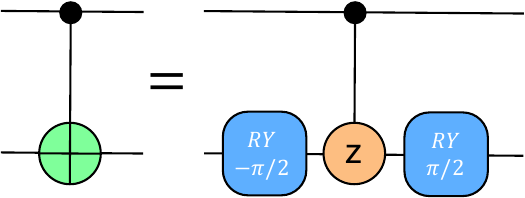}
  \caption{Demonstration of the compilation of a \textsc{cnot} gate into a CZ gate, flanked by two RY gates.}
  \label{fig_compileCX}
\end{figure}

We basically applied QPT to two fundamental two-qubit gates based on the superconducting quantum computing platform: the controlled-Z (CZ) gate and the controlled-phase (CP) gate. It is important to note that other composite gates, such as the controlled-\textsc{not} (\textsc{cnot}) gate, can be decomposed into these basic two-qubit gates in conjunction with additional single-qubit gates, like in Fig.~\!\ref{fig_compileCX}. For any two-qubit CZ gates we use, we characterized them independently with QPT.

The flexibility of our framework allows for the characterization of arbitrary processes, albeit with increased computational costs. In the context of practical quantum computing experiments, single-qubit gates typically exhibit high fidelity, making errors in two-qubit gates more significant. In our numerical simulation with real noise, which is called QPT-MPDO simulation, we performed the single-qubit gates as ideal gates with consideration of practical experimental conditions and efficiency. However, if single-qubit gate errors are deemed non-negligible, they can be included in the simulations to provide a more accurate assessment.

The final numerical simulation accuracy will also be affected by the accuracy of QPT, especially when we sacrifice the accuracy of QPT in exchange for efficiency. Recent advancements, such as those leveraging machine learning, offer more efficient characterization of quantum circuits and can achieve good approximation effects with high efficiency, as referenced in the literature~\cite{torlai_quantum_2023}. This is especially relevant in realistic NISQ scenarios, where experimental platform errors are more complex than simple noise models like depolarization. Generally, the QPT of two or three qubits is experimentally straightforward and not significantly affected by computational complexity. For more complex error scenarios, QPT can be performed on multiple layers of quantum gates or multiple qubits to form one ``quantum gate'' for the circuit. In these cases, less accurate QPT methods might be necessary to maintain efficiency. For typical experimental scenarios, we believe that the potential accuracy deviation from QPT should fall within the error range of repeated experiments, especially for parameterized quantum approximate optimization algorithm (QAOA) scheme. However, it is crucial to note that the performance of experimental platforms can change dynamically. For accurate evaluation of the experimental scheme at any given time, the QPT data should be as current as possible.

\section{Simulations \& Applications}
\label{sec4}
\subsection{Variational Approach to Entangled States Generation}
We assess the performance of our simulator under real operational conditions by employing the QAOA to generate an entangled quantum state. This application tests the simulator with a variational quantum circuit and offers valuable insights into the behavior of quantum circuits amidst diverse environmental and systemic disruptions commonly encountered in real quantum computing settings.

We employ $2N$ measurement qubits to probe the state of a target qubit. The simulated Hamiltonian for these $2N{+}1$ qubits follows an Ising model, where the target qubit is subjected to a transverse field Ising Hamiltonian, described as
\begin{equation}
    H(t) = -\sum_{m} J_{sm}(t){\sigma}_z^s{\sigma}_z^m - \!\!\sum_{m,n\neq s}\!\!{J}_{mn}{\sigma}_z^m{\sigma}_z^n - h(t)\sum_m{\sigma}_x^m,
  \label{eq_hamiltonian}
\end{equation}
where $J$ represents the coupling of ZZ interaction and $h(t)$ is the transverse field applied to the measured qubits. In the process of numerical simulation, the strength of the ZZ interaction is set to $J_{sm} = 1$.

\begin{figure}[htbp]
  \centering
  \includegraphics[width=\linewidth]{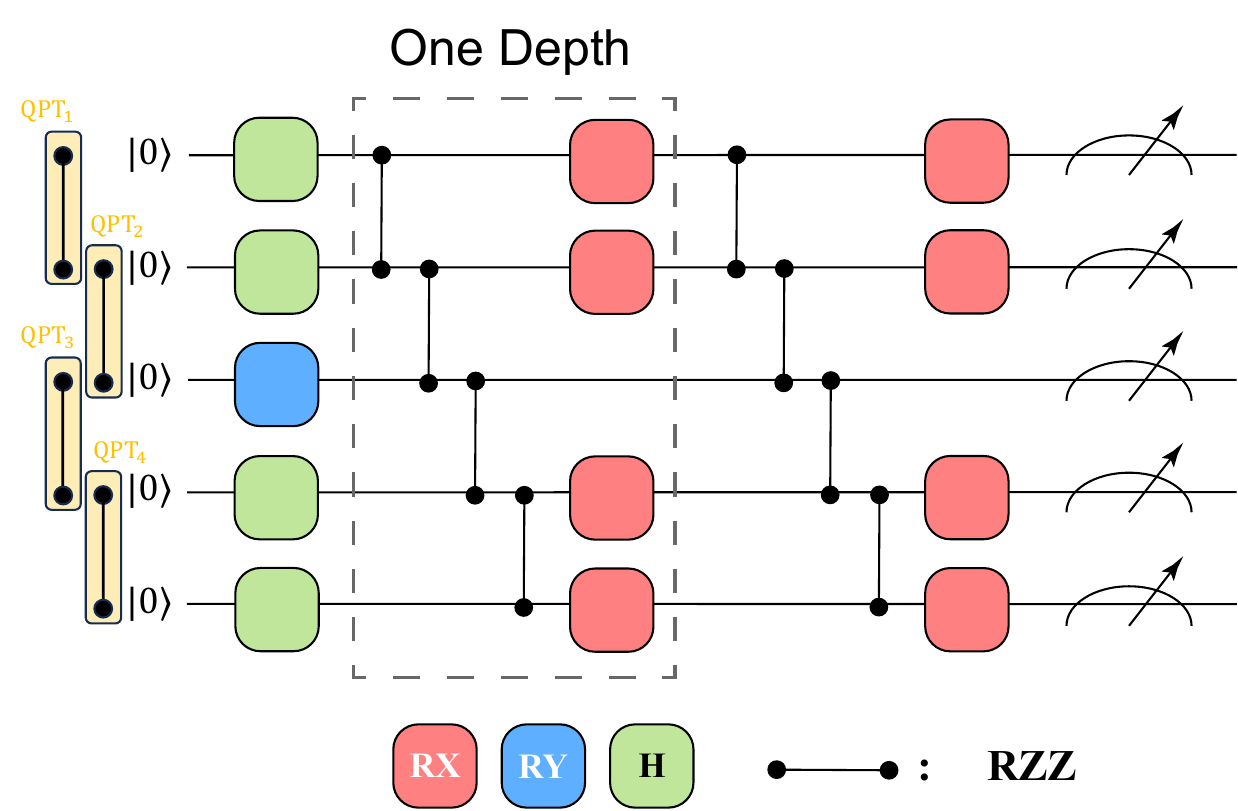}
  \caption{Schematic of five-qubit QAOA circuit with nearest-neighbor interaction. The RZZ gates symbolize the ZZ interaction. This circuit design is scalable to larger qubit systems while maintaining the same structural framework. It is usually better to choose a circuit depth equal to half the number of qubits. Each independent qubit-interval CZ gate is individually characterized by QPT before operation.}
  \label{fig_exp2_circ}
\end{figure}

\begin{figure}[htbp]
  \centering
  \includegraphics[width=\linewidth]{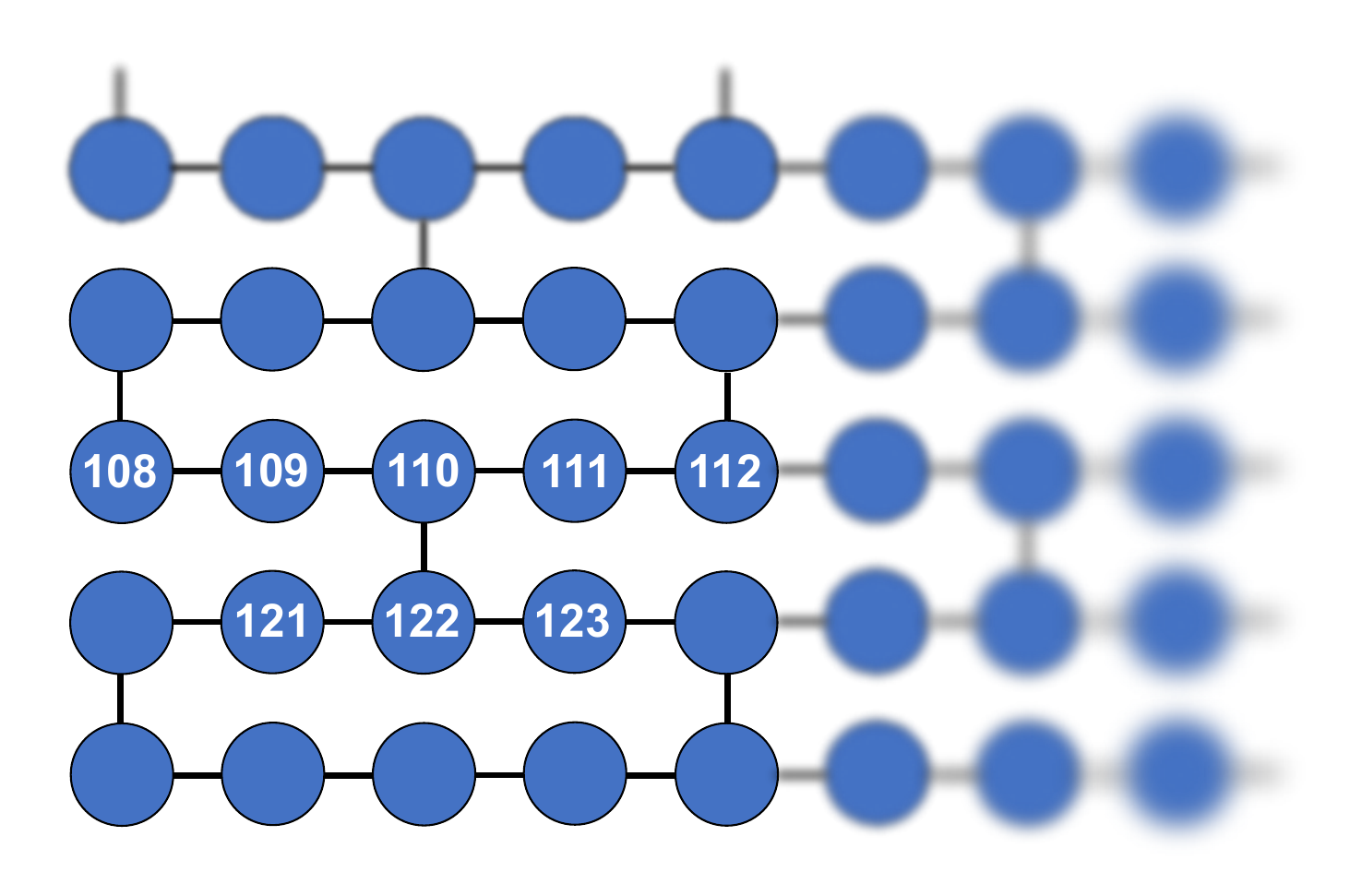}
  \caption{Part of topology graph from Baiwang device in the Quafu cloud platform.}
  \label{fig_chip_info}
\end{figure}

The typical QAOA circuit is shown in Fig.~\ref{fig_exp2_circ}, which includes the RZZ gate $RZZ(\gamma)\equiv\mathrm{exp}({\mathrm{i} \gamma\hat{\sigma}_z\hat{\sigma}_z})$, the RX gate $RX(\theta)\equiv\mathrm{exp}({\mathrm{i} \theta\hat{\sigma}_x})$, and the RY gate $RY(\theta)\equiv\mathrm{exp}({\mathrm{i} \theta\hat{\sigma}_y})$. This design is scalable and maintains consistent structure across larger qubit systems, with the circuit depth ideally being half the number of measurement qubits. To highlight the impact of nontrivial noise on real quantum computers, we perform the 5-qubit demonstration by using the \textit{Baiwang} device on the Quafu cloud platform~\cite{quafu_website}. A partial structure of this device is shown in Fig.~\!\ref{fig_chip_info} and its basic information is listed in Table~\ref{sheet_1}.

\begin{table}[htbp]
  \begin{tabular}{lcccccccccc}
    \toprule
    \toprule
    Qubit &
      \multicolumn{2}{c}{$Q_{108}$} &
      \multicolumn{2}{c}{$Q_{109}$} &
      \multicolumn{2}{c}{$Q_{110}$} &
      \multicolumn{2}{c}{$Q_{111}$} &
      \multicolumn{2}{c}{$Q_{112}$} \\
    \midrule
    $T_1$($\mu s$) &
      \multicolumn{2}{c}{$17.01$} &
      \multicolumn{2}{c}{$42.25$} &
      \multicolumn{2}{c}{$26.48$} &
      \multicolumn{2}{c}{$38.03$} &
      \multicolumn{2}{c}{$22.53$} \\
    $T_2$($\mu s$) &
      \multicolumn{2}{c}{$22.58$} &
      \multicolumn{2}{c}{$26.36$} &
      \multicolumn{2}{c}{$21.57$} &
      \multicolumn{2}{c}{$14.85$} &
      \multicolumn{2}{c}{$21.77$} \\
    Qubit frequency~\!(GHz) &
      \multicolumn{2}{c}{$4.294$} &
      \multicolumn{2}{c}{$4.550$} &
      \multicolumn{2}{c}{$4.267$} &
      \multicolumn{2}{c}{$4.350$} &
      \multicolumn{2}{c}{$4.413$} \\
    Mean readout fidelity &
      \multicolumn{2}{c}{$0.958$~} &
      \multicolumn{2}{c}{$0.977$~} &
      \multicolumn{2}{c}{$0.985$~} &
      \multicolumn{2}{c}{$0.963$~} &
      \multicolumn{2}{c}{$0.978$} \\
    CZ gate error rate &
      {~~~} &
      \multicolumn{2}{c}{$0.957$~} &
      \multicolumn{2}{c}{$0.970$~} &
      \multicolumn{2}{c}{$0.980$~} &
      \multicolumn{2}{c}{$0.950$} &
      {} \\
      \bottomrule
      \bottomrule
  \end{tabular}
  \caption{Basic information of physical qubits on the Baiwang device from the Quafu cloud platform, including CZ gate error rates between physically connected neighboring qubits, $T_1$, $T_2$ and frequency of qubits.}
  \label{sheet_1}
\end{table}

\begin{figure}[htbp]
  \centering
  \includegraphics[width=\linewidth]{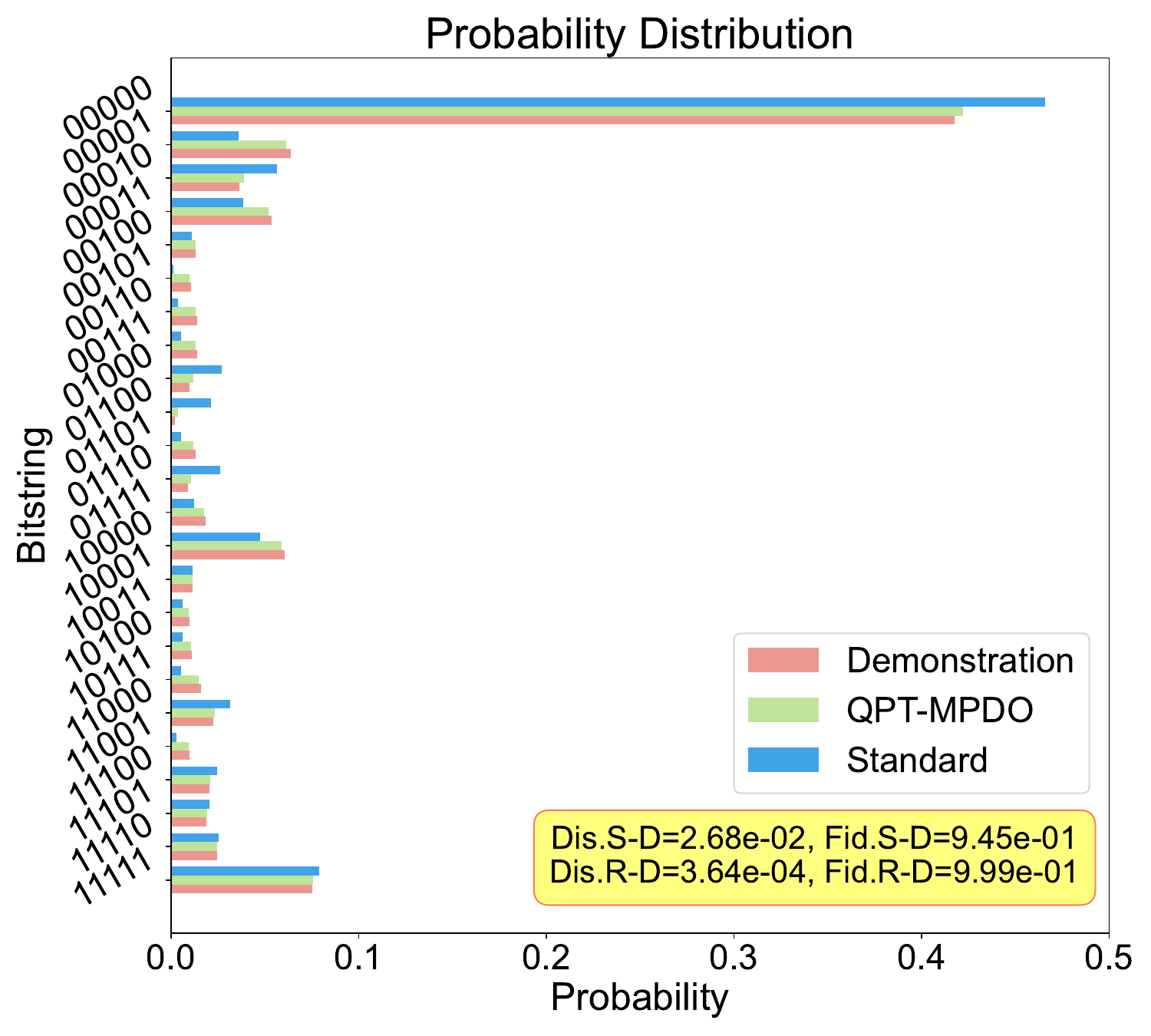}
  \caption{Probability distributions from standard and QPT-MPDO simulations compared with demonstration result, utilizing Jensen-Shannon divergence for distribution distance measurement. Orange bars represent demonstration results from the Quafu cloud platform, blue bars represent simulations with a standard quantum noise model parameterized near physical parameters, and green bars show simulations using real noise QPT data from Quafu. The standard noise model inadequately mimics the actual system, showing a divergence of $2.68*{10}^{-2}$, while the real noise model aligns more closely with demonstration outcomes with divergence of $3.64*{10}^{-4}$. Minor discrepancies between QPT-MPDO simulation and demonstration data may arise due to the truncation processes in the tensor network framework.}
  \label{fig_exp2}
\end{figure}

Here we take the initial state of the target qubit as $\ket{\psi_0} = \sin({\pi}/{8})\ket{0} + \cos({\pi}/{8})\ket{1}$ for demonstration. Executing the quantum circuit yielded results depicted in Fig.~\!\ref{fig_exp2}, with pink bars representing demonstration data from Quafu and blue bars indicating numerical simulation outcomes using the standard noisy circuit model, which is described in Fig.~\!\ref{fig_unifiedNoise}(c), including depolarization and thermal relaxation noise. To assess quantum state similarity, we employed quantum state fidelity $F_{\psi, \varphi} = \left|{\bra{\psi}\ket{\varphi}}\right|^2$, and the Jensen-Shannon divergence
\begin{equation}
  \mathrm{JSD}(P||Q) = \frac{1}{2}\left(
    \sum_i P_i\ln\frac{P_i}{M_i} + \sum_i Q_i\ln\frac{Q_i}{M_i}
  \right),
\end{equation}
where $P$ and $Q$ denote two discrete probability distributions, and each $P_i$ and $Q_i$ denotes the $i$-th element of corresponding distributions, respectively. The average distribution $M$ is defined as the midpoint between $P$ and $Q$, calculated as $M = \frac{1}{2}(P + Q)$. 

We attempted to incorporate the physical qubit information into a quantum circuit model with standard noise, and further used gradient descent (with Jensen-Shannon divergence as the loss function) to optimize the parameters to find a set of `optimal' parameters that are closest to the demonstration data. The improved numerical simulation accuracy, however, still surpassed the performance of the real device, suggesting that simplistic noise models fail to encapsulate the entirety of the real noise profile.

By utilizing QPT to obtain the $\chi$ matrix data of CZ gates, which can be converted into a Kraus operator as a noisy quantum gate by the process in Fig.~\!\ref{fig_chi2kraus}, for the qubits in Table~\ref{sheet_1} from Quafu, we are able to enhance our numerical simulation to closely replicate the demonstration. This is evident in the improved alignment between the result of QPT-MPDO simulation (green bars in Fig.~\ref{fig_exp2}) and the demonstration data, emphasizing the importance of integrating comprehensive noise considerations into quantum simulations. 

\begin{figure*}[htbp]
  \centering
  \includegraphics[width=0.85\linewidth]{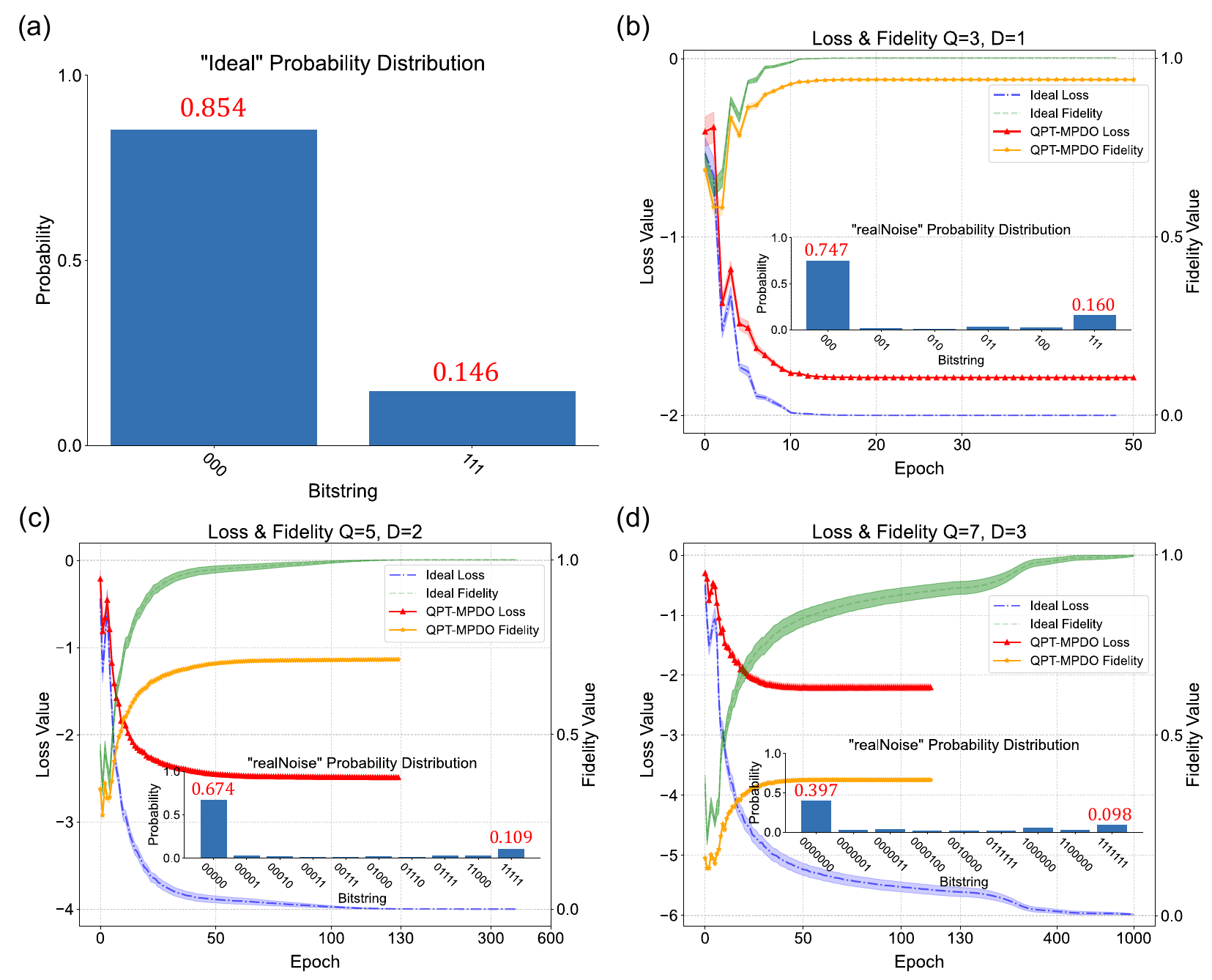}
     \caption{Simulation results for the entangled quantum state generation using a QAOA circuit across varying qubit configurations. (a) Probability distribution for an ideal simulation with three qubits, establishing a reference case for this QAOA application for initial state $\ket{\psi_0} = \sin(\pi/8)\ket{0} + \cos(\pi/8)\ket{1}$. (b-d) Comparative analysis of ideal and noisy simulations for three-qubit and one-depth (b), five-qubit (c), and seven-qubit (d) systems. The main graphs track the loss function and state fidelity across simulation epochs, employing a gradient descent algorithm, with the shaded areas representing error bars. The insets highlight the respective bitstring probability distributions, reflecting the distinction between ideal and real-noise conditions.}
  \label{fig_exp2_num}
\end{figure*}

After verifying the capability of our simulator to effectively handle real noise in contrast to standard models, we next explore the performance of QAOA on a larger array of qubits and deeper quantum circuits with real noise data. The influence of real quantum noise on quantum states is detailed in Fig.~\!\ref{fig_exp2_num}. We investigate the effectiveness of QAOA across different circuit complexities, as shown by our numerical simulation for various qubit configurations.

Figure~\ref{fig_exp2_num}(a) shows the probability distribution of bitstrings from an ideal three-qubit simulation, establishing a baseline for this QAOA circuit configuration. Figures~\ref{fig_exp2_num}(b)-\ref{fig_exp2_num}(d) display the results from both ideal and noisy conditions, tracking the evolution of the loss function and quantum state fidelity over multiple epochs. The loss function indicates the progression of the optimization via the gradient descent algorithm, while the fidelity measures how closely the generated state matches the desired entangled state. These metrics are evaluated for systems with three, five, and seven qubits.

The shaded areas around the loss and fidelity curves represent error bars. These results confirm the ability of algorithm to approximate the entangled quantum state under ideal conditions. However, they also reveal the challenges posed by noise, highlighted by the differences between the ideal and noisy simulations. This underscores the significant impact of real quantum noise on quantum state preparation, emphasizing the importance of robust simulation practices in real-world quantum computing scenarios.

\subsection{Noisy MaxCut Problems}
\begin{figure}[H]
  \centering
  \includegraphics[width=0.8\linewidth]{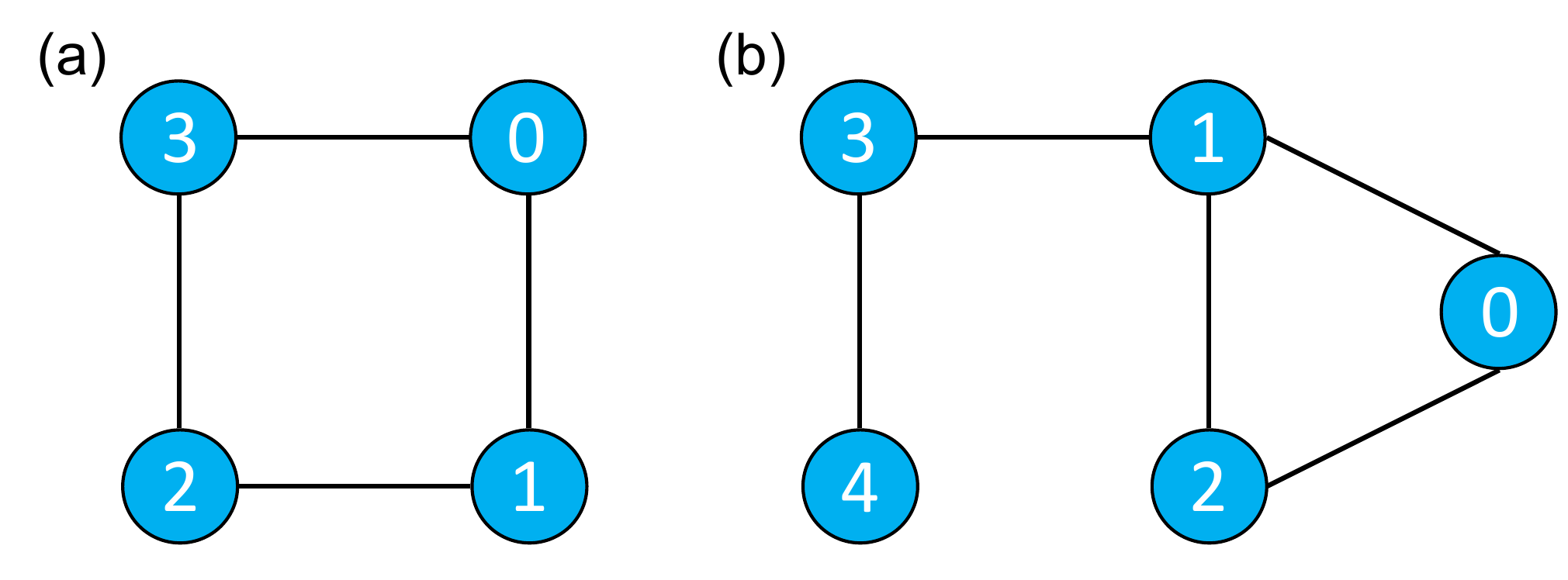}
  \caption{Two example graphs for MaxCut problem.}
  \label{fig_exp3_I}
\end{figure}

Beyond simulating simple one-dimensional (1D) systems, fully-connected simulations can be conducted, exemplified by solving the MaxCut problem in mathematics using the QAOA~\cite{9826192}. Figure~\!\ref{fig_exp3_I} shows two MaxCut scenarios: One with a graph of four qubits (vertices) and another with five. The objective of the MaxCut problem is to maximize the number of edges that are severed by partitioning the vertices of the graph into two distinct subsets~\cite{10.5555/574848, 10.1145/227683.227684}.

\begin{figure*}[t]
  \centering
  \includegraphics[width=\linewidth]{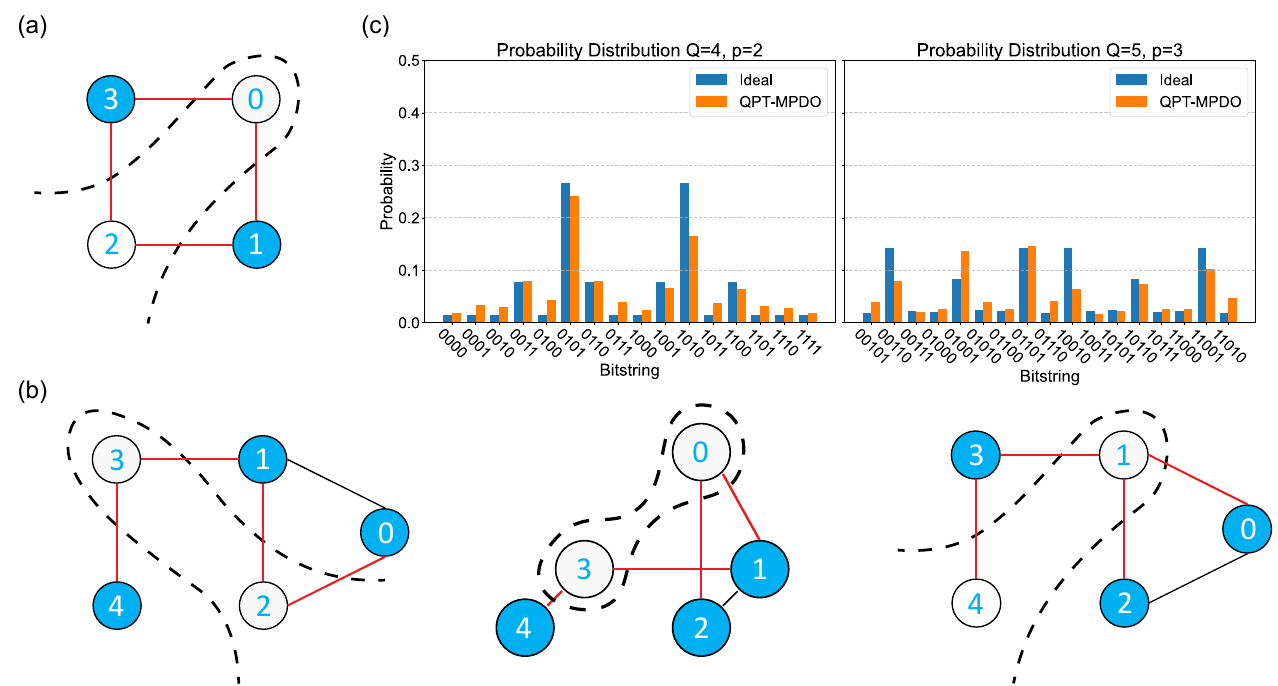}
  \caption{Maximum cut problem on full-connection simulator. (a) Illustration of a four-qubit graph where the optimal partition \{(0, 2), (1, 3)\} results in all four edges being cut; (b)  For a five-qubit graph, three optimal partitioning methods—\{(2, 3), (0, 1, 4)\}, \{(0, 3), (1, 2, 4)\}, and \{(1, 4), (0, 2, 3)\}—achieve a maximum of four-edge cuts; (c) Outcome of the Quantum Approximate Optimization Algorithm applied to the maximum cut problem, with results for qubits number $Q=4$ and circuit depth $p=2$ leading to quantum states $\ket{0101}$ and $\ket{1010}$, indicating the grouping of qubits (vertices) into sets. For $Q=5$ and $p=3$, the quantum states $\ket{00110}$, $\ket{01001}$, $\ket{01101}$, $\ket{11001}$, $\ket{10110}$ and $\ket{10010}$ represent the corresponding vertex classifications as detailed in panel (b). The blue and orange bars denote the simulation outcomes with ideal quantum gates and with real gate noise, respectively.}
  \label{fig_exp3}
\end{figure*}

In the general case, consider a graph with $m$ edges and $n$ vertices. We seek a partition $z$ that divides the vertices into two sets, i.e. $A$ and $B$, to maximize the cut count
\begin{equation}
C(z) = \sum_{\gamma = 1}^m C_\gamma(z),
\label{eq_maxcut}
\end{equation}
where $C$ represents the total number of edges cut by the partition $z$. For each edge $\gamma$, if the partition $z$ places one vertex in set $A$ and the other in set $B$, then $C_\gamma(z)=1$; otherwise, $C_\gamma(z)=0$. To cast this into a quantum framework, we represent vertices as qubits and encode the measured bitstring $z=z_1z_2\dots z_n$ as the partitioning outcome, where $z_j = 0$ implies that the $j$-th qubit is in set $A$, and $z_j = 1$ indicates it is in set $B$.

For the $\gamma$-th edge connecting vertices $i$ and $k$, the objective function is expressed as
\begin{equation}
C_\alpha = \frac{1}{2}\left(1 - \langle{\sigma_z^i\sigma_z^k}\rangle\right),
\label{eq_maxcut_obj}
\end{equation}
where $\sigma_z$ denotes the Pauli-Z operator. If the qubits $Q_i$ and $Q_k$ are in different partitions, $C_\alpha$ attains an eigenvalue of $1$, fulfilling the MaxCut problem's criterion.

To implement the QAOA, the quantum circuit initiates with a layer of Hadamard gates to create a uniform superposition across the $n$ bitstring basis states. In a standard QAOA circuit, parameterized quantum gates are employed, specifically $\mathrm{R\mathrm{X}}(\alpha)$ and $\mathrm{R{\mathrm{ZZ}}}(\beta)$ gates in this case, defined as
\begin{equation}
  \begin{aligned}
    &U_{\mathrm{RX}_l} = \prod_{i=1}^n \mathrm{e}^{-\mathrm{i}\alpha_l\sigma_x^i},\\
    &U_{\mathrm{RZZ}_l} = \prod_{E_{i, k}} \mathrm{e}^{-\mathrm{i}\beta_l(1-\sigma_z^i\sigma_z^k)/2},
  \end{aligned}
\end{equation}
where $E_{i, k}$ denotes the edge between the qubits $Q_i$ and $Q_k$, and $l$ signifies the $l$-th depth in the circuit, which is composed of one layer of $U_{\mathrm{RX}_l}$ and $U_{\mathrm{RZZ}_l}$.

For a circuit with $p$ depths, the sequence of gate operations is given by
\begin{equation}
\ket{\psi_{\vec{\alpha}, \vec{\beta}}} = U_{\mathrm{RX}_p}U_{\mathrm{RZZ}_p}\cdots U_{\mathrm{RX}_1}U_{\mathrm{RZZ}_1}H^n\ket{0^n},
\end{equation}

where $\vec{\alpha} = (\alpha_1, \ldots, \alpha_p)$ and $\vec{\beta} = (\beta_1, \ldots, \beta_p)$ represent the sets of parameters, and $H$ denotes the Hadamard gate. The goal is to optimize these parameters so that the expectation value of the final state $\ket{\psi_{\vec{\alpha}, \vec{\beta}}}$ with respect to the operator defined in Eq.~\!(\ref{eq_maxcut}) is maximized by minimizing the $-C$, representing the objective function for the MaxCut problem.

Upon executing the quantum circuit for the graphs depicted in Fig.~\!\ref{fig_exp3_I}, we identify the quantum states that optimize the objective function, along with their respective MaxCut configurations. For the case illustrated by the left image of Fig.~\!\ref{fig_exp3}(c) with qubits number $Q=4$ and circuit depth $p=2$, the states $\ket{0101}$ and $\ket{1010}$ emerge with high probability. These states correspond to the MaxCut depicted in~\!\ref{fig_exp3}(a), where the qubits $Q_0$ and $Q_2$ are segregated into set $A$, while the remaining qubits fall into set $B$, delineated by a black dashed line. This grouping results in the cutting of four edges, achieving the maximum cut possible in this graph. Notably, the quantum states $\ket{0101}$ and $\ket{1010}$ produce identical partitioning outcomes, serving as mirror images of each other. Similarly, for the right image in Fig.~\!\ref{fig_exp3}(c) with $Q=5$ and $p=3$, the states $\ket{00110}$, $\ket{01001}$, $\ket{01101}$, $\ket{11001}$, $\ket{10110}$, and $\ket{10010}$ manifest with significant probability, mirroring the scenario in Figs.~\!\ref{fig_exp3}(a) and \ref{fig_exp3}(b). These six states correspond to three distinct MaxCut configurations, as depicted in Fig.~\!\ref{fig_exp3}(b), each achieving the maximum cut by severing $4$ edges.

In Fig.~\!\ref{fig_exp3}(c), the blue bars represent the idealized, noiseless circuit outcomes, whereas the orange bars depict the results under the influence of real noise encountered. Note that the introduction of noise conspicuously reduces the probabilities of states with a `high' bitstring composition. The decrease in probability is particularly pronounced in states with higher probabilities in ideal conditions, highlighting the degradation effects of noise on the performance of the system. The results validate the expectation that noise introduces significant perturbations in state probabilities, underscoring the challenge of achieving high-fidelity quantum operations in practical quantum computing environments.

\subsection{Effects of Noise Truncation}
\begin{figure}[H]
  \includegraphics[scale=0.46]{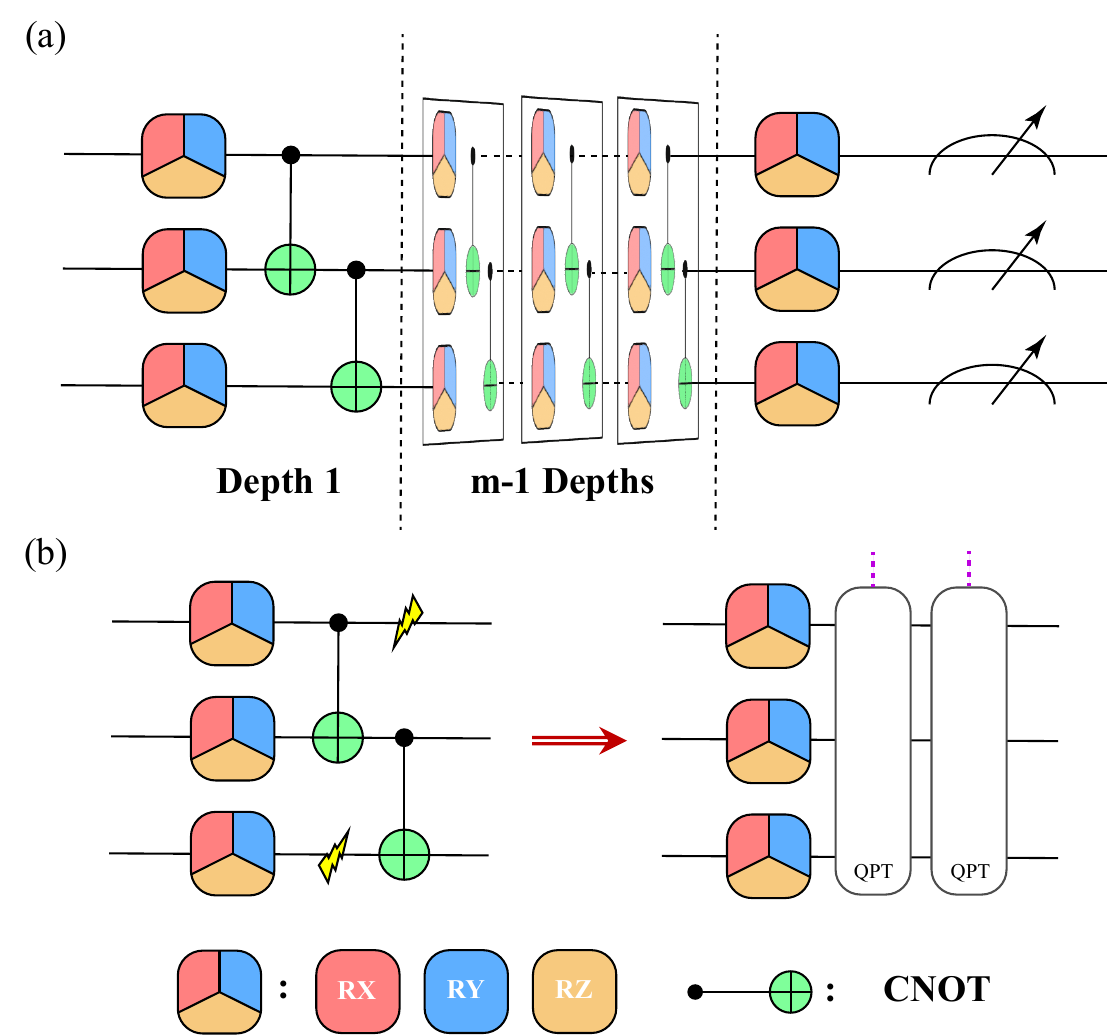}
  \captionsetup{justification=raggedright,singlelinecheck=false}
  \caption{Overview of quantum circuit composition and crosstalk handling. (a) Illustration of a randomized quantum circuit for three qubits (denoted as $Q=3$). Each layer of the circuit comprises a set of single-qubit gates—randomly chosen from $\mathrm{RX}$, $\mathrm{RY}$, and $\mathrm{RZ}$ gates—followed by a layer of \textsc{cnot} gates representing entanglement operations. (b) Integration of a two-qubit gate into the simulation circuit, represented as a tensor, incorporates the effects of crosstalk by including interactions with the adjacent qubit. The bottom legend identifies the gate symbols used in the circuit diagram. It is readily applicable to experimental data; for instance, as shown in (b), one simply performs QPT on all qubits deemed to be influenced by noise and integrates this data into the simulation.}
  \label{randomCirc_fig}
\end{figure}

To assess the resilience of QPT-MPDO simulations against dimensionality truncation, we employ randomly generated one-dimensional quantum circuits. Our objective is to ascertain the critical truncation threshold where the fidelity of the simulated system markedly diverges from that of the actual quantum system, as depicted in Fig.~\!\ref{randomCirc_fig}. The single-qubit gates of the circuit, represented by three-colored boxes, are randomly chosen from the set ${RX(\alpha), RY(\beta), RZ(\gamma)}$, with the parameters $\alpha$, $\beta$, and $\gamma$ uniformly sampled from the range $(0, 2\pi]$. This ensures a thorough exploration of the Hilbert space. Following the single-qubit gate layer, a layer of \textsc{cnot} gates is applied to introduce entanglement, thereby constructing the circuit's depth composed of single- and two-qubit gates. The iterative application of these layers, $m-1$ times, is indicated by the two dotted lines. A final layer of single-qubit gates precedes the measurement phase to augment the randomness of the system.

In Fig.~\!\ref{randomCirc_fig}(b), we illustrate the crosstalk effects resulting from two-qubit gate operations on adjacent qubits. To model these effects, we introduce an $RZ(\alpha)$ gate with a small, random strength onto the nearest neighbor qubit, depicted with a lightning bolt icon. Generally, crosstalk intensity is minimal in a faultless experimental setup, we can estimate the angle of the $\mathrm{RZ}$ gate using the methods described in~\cite{PRXQuantum.3.020301, PhysRevB.104.045420}. For this simulation, the angle $\alpha$ was arbitrarily chosen within the range $(10^{-5}\pi, 10^{-3}\pi]$ to reflect the potential variability in a real-world experiment.

To effectively assess the density matrix overlap before and after real noise reduction, it is common practice to employ quantum fidelity as a descriptor~\cite{doi:10.1080/09500349414552171}

\begin{equation}
  F(\rho, \sigma) = \left(\mathrm{tr} \sqrt{\sqrt{\rho}\sigma\sqrt{\rho}}\right)^2.
  \label{eq_fidelity}
\end{equation}

\begin{figure*}[htbp]
  \centering
  \includegraphics[width=0.9\linewidth]{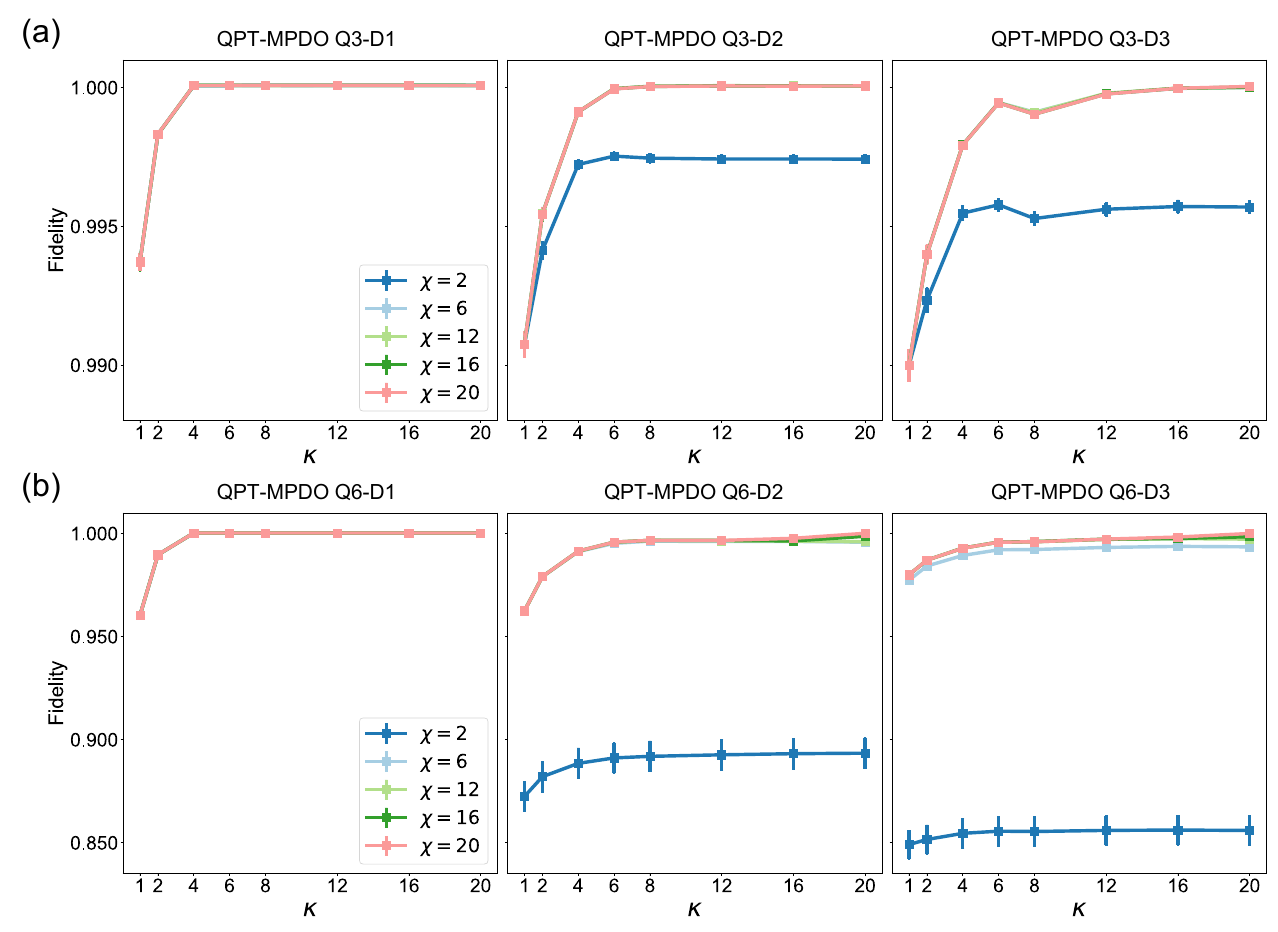}
  \caption{Fidelity of truncated quantum states under varying maximum bond dimension $\chi$ and maximum inner dimension $\kappa$ along with error bars, calculated from 1,000 random circuits for two scenarios: (a) qubit count $Q=3$ and (b) qubit count $Q=6$. For smaller number of quantum bits or shallower quantum circuit depths, employing smaller $\chi$ and $\kappa$ values enables highly faithful numerical simulation of real noise in quantum circuits. As the circuit depth and system size increase, greater noise and entanglement are introduced into the quantum system. Consequently, it becomes necessary to scale up the values of $\chi$ and $\kappa$ proportionally to achieve sufficiently accurate approximations for simulating the scalable noisy system.}
  \label{exp1_fig}
\end{figure*}

Referring to Eq.~\!(\ref{eq_fidelity}), we subject $\sigma$ to dimensional cropping under various $\chi$ and $\kappa$ values and $\rho$ is the experimental real noisy density matrix without truncation. Figure~\ref{exp1_fig} clearly illustrates that, in the $Q=3$ circuit series, fidelity remains relatively stable across different depths, indicating that truncation has a minor information loss. Conversely, in the $Q=6$ series, the impact of truncation is more pronounced, especially as depth increases, evidenced by the more significant drop in fidelity with lower $\chi$ and $\kappa$ values. This suggests that larger systems may require higher bond dimensions to maintain fidelity, particularly for deeper circuits. However, compared to the full dimension, they are still small enough for a classic computer to simulate. 

Overall, the result indicates that the choice of bond dimension for dimension truncation is a critical factor in maintaining fidelity in noisy simulations, and this choice becomes more crucial as the system size and circuit complexity increase. The cost of simulation approx to $ND\chi^3\kappa^3$~\cite{cheng_simulating_2021}, where $N$ and $D$ represent the number of qubits and the depth of circuit respectively. It is tricky whether an idealized noise is easier to be simulated in MPDO framework. However, it can be expected that, in general, a model with real noise requires a noise bond dimension $\kappa$ that is greater than or equal to that of an idealized error model. Real noise models may have more complex principal components, necessitating a higher noise bond dimension to achieve the same truncation fidelity as idealized error models. This depends on the complexity of the actual quantum process.

\section{Conclusion \& Discussion}
\label{sec5}
In conclusion, we introduce a MPDO-based noise simulator for the numerical simulations of noisy quantum circuits, focusing on circuits with QPT data measured by real experiments. This simulator allows for a systematic analysis of the effects of noise truncation, highlighting the importance of bond dimension selection in preserving simulation fidelity, especially as the system size and circuit complexity increase. Our approach is exemplified in the study of the QAOA for preparing entangled states and solving MaxCut problems under noisy conditions, offering insights into the operational dynamics of quantum algorithms in realistic settings.

Furthermore, our research could facilitate the pre-experimental testing of quantum schemes, enabling researchers to evaluate the viability of experimental setups before actual implementation with less effort. This predictive capability is crucial for designing experiments and optimizing resources, thus saving time and reducing the likelihood of costly errors. The use of efficient QPT methods in our simulator enhances the capture of experimental noise characteristics, thereby improving the accuracy and realism of our simulations, thus also making our approach an efficient tool for designing noise-resilient quantum circuits aiming at NISQ algorithms. In fact, experimental platforms will experience instability as time passes after calibration, the need for up-to-date QPT data becomes critical to ensure accurate assessments and optimizations.

A significant aspect of our work is its potential application to error-mitigation strategies. The flexibility of the MPDO approach allows for the evaluation of the effectiveness of quantum circuits under various noise conditions, with its noise bond formally representing the presence of quantum noise. This approach enables researchers to deliberately design experimental noise scenarios, facilitating the investigation and mitigation of specific quantum noise encountered in actual experiments. Furthermore, methods discussed in Ref.~\cite{guo_quantum_2022}, such as using MPO to calculate the inverse of noisy quantum processes, can be integrated into our framework, offering pathways for mitigating unknown quantum noise.

Moreover, our approach addresses both Markovian and non-Markovian noise. By characterizing quantum processes at different time or space scales—whether single-qubit gates, multiqubit gates, or multilayer quantum gates—we can model various noise interactions with appropriate levels of detail. This capability is essential for understanding the full impact of noise on quantum computations and developing effective mitigation strategies.

Our study underscores the importance of considering complex noise profiles in practical quantum computing settings. Advanced simulation techniques, like those we developed, are crucial for evaluating the feasibility and performance of NISQ devices. Future research could further explore integrating these techniques with specific error-mitigation strategies, potentially correcting sources of error in experimental processes and extending the utility of our simulator in designing robust quantum circuits. In summary, our MPDO-based simulator not only advances the understanding of noise impacts on quantum circuits but also provides a versatile framework for enhancing the fidelity and robustness of quantum computations in practical noisy environments.

\begin{acknowledgments}
  This work was supported by National Natural Science Foundation of China (Grants No. 92265207, No. T2121001, No. 11934018, No. 12122504, and No. T2322030), Beijing Natural Science Foundation (Grant No. Z200009), the Innovation Program for Quantum Science and Technology (Grant No. 2021ZD0301800), Beijing Nova Program (Grants No. 20220484121 and No. 2022000216), and China Postdoctoral Science Foundation (Grant No. GZB20240815).
\end{acknowledgments}

\providecommand{\noopsort}[1]{}\providecommand{\singleletter}[1]{#1}%

\end{document}